\documentclass[12pt,a4wide]{article}
\usepackage{amsmath,amssymb,amsfonts}

\newcommand{\eqa}{\begin{eqnarray}}
\newcommand{\ena}{\end{eqnarray}}
\newcommand{\topstar}[1]{\setlength{\unitlength}{1mm}
\begin{picture}(2,0)(-1,-1.4)
   \put(0,0){\makebox(0,0){$#1$}}
   \put(0,2.4){\makebox(0,0){\mbox{\tiny$\star$}}}
\end{picture}}
\usepackage{epsfig}
\begin{document}
\begin{center}
{\large {\bf Classical Electrodynamics in Quasi-Metric Space-Time}}
\end{center}
\begin{center}
Dag {\O}stvang \\
{\em Department of Physics, Norwegian University of Science and Technology 
(NTNU) \\ N-7491 Trondheim, Norway}
\end{center}
\begin{abstract}
The quasi-metric manifold $\cal N$ is equipped with two 
one-parameter families of metric tensors ${\bf {\bar g}}_t$ and ${\bf g}_t$,
each parametrized by the global time function $t$. Moreover, in
$({\cal N},{\bf {\bar g}}_t)$, one must define two different electromagnetic 
field tensor families corresponding to the active field tensor family
${\bf {\tilde F}}_t$ and the passive field tensor family ${\bf {\bar F}}_t$, 
respectively. The active electromagnetic field tensor family 
${\bf {\tilde F}}_t$ couples to gravity. By construction, the norm of the 
passive electromagnetic field tensor family ${\bf {\bar F}}_t$ experiences
a secular decrease, defining a global cosmic attenuation (not noticeable 
locally) of the electromagnetic field. Local conservation laws for passive 
electromagnetism imply that ${\bf {\bar {\nabla}}}{\cdot}{\bf {\bar F}}_t=0$ in
electrovacuum, ensuring that photons move on null geodesics of 
$({\cal N},{\bf {\bar g}}_t)$. From ${\bf {\bar F}}_t$, one may construct the 
passive electromagnetic field tensor family ${\bf F}_t$ in 
$({\cal N},{\bf g}_t)$ in the same way as ${\bf g}_t$ is constructed from 
${\bf {\bar g}}_t$. This ensures that photons move on null geodesics of 
$({\cal N},{\bf g}_t)$ as well. As a simple example, the (exact) quasi-metric 
counterpart to the Reissner-Nordstr\"{o}m solution in General Relativity is 
calculated. Besides, it is found that a classical charged test particle 
electromagnetically bound to a central charge will participate in the cosmic 
expansion. But since quantum-mechanical states should be unaffected by the 
expansion, this classic calculation is hardly relevant for quantum-mechanical 
systems such as atoms, so there is no reason to think that the cosmic expansion
should apply to them. Finally, it is shown that the main results of geometric 
optics hold in quasi-metric space-time.
\end{abstract}
\topmargin 0pt
\oddsidemargin 5mm
\renewcommand{\thefootnote}{\fnsymbol{footnote}}
\section{Introduction}
Recently, the so-called quasi-metric framework (QMF) as a geometric basis for 
relativistic gravity [1], was introduced as a possible alternative to the usual
metric framework (MF) underlying metric theories of gravity. The QMF is similar
to the MF in some respects (e.g., both are based on the Einstein equivalence 
principle), but one aspect of the QMF having no counterpart in the MF is the
existence of a non-metric sector. Another characteristic property of the QMF
is that it necessarily exists two fundamentally different types of dimensional 
scales (i.e., length scales, time scales, mass scales) in the quasi-metric
universe; one gravitational and one atomic. The atomic scale can be constructed
operationally from physical systems where gravitational interactions are 
insignificant. In practice, this means that local measurement devices based on 
non-gravitational physics operationally define ``atomic units'' and that both 
gravitational and non-gravitational scales are expressed in terms of such 
units. Now the QMF predicts that expressed in atomic units, there should exist 
systematic scale changes between gravitational and non-gravitational systems. 
Moreover, one manifestation of said non-metric sector is the {\em global} part 
of these scale changes, which is identified with the global cosmic expansion. 
Thus, within the QMF, the cosmic expansion is interpreted differently and has 
a different mathematical description than within the MF. And as a consequence, 
the cosmic expansion is predicted to affect gravitational systems regardless 
of scale. This prediction has a number of observable consequences, some of 
which have been detected in the solar system; see refs. [2, 3] for details. 
Furthermore, how matter fields couple to gravity cannot be independent of said 
systematic scale changes. Therefore, the QMF does not fulfil the strong 
equivalence principle since it is necessary to separate between {\em active 
mass-energy} as a source of gravitation and {\em passive mass-energy} entering 
the equations of motion. 

No exceptions from these considerations are made for the electromagnetic field;
in particular, it is necessary to separate between the active and passive 
aspects of electromagnetic mass-energy. This means that a description of 
classical electrodynamics within the QMF must differ from its counterpart 
within the MF. It turns out that there are three crucial differences. Firstly, 
global scale changes between electromagnetism and gravitation should be present
explicitly as a general time dependence in the norm of the electromagnetic 
field tensor. By construction, this general time dependence takes the form of a
global cosmic ``attenuation'' of the electromagnetic field. Secondly, 
the necessity to separate between active and passive aspects of 
electromagnetism involves the introduction of {\em two} fundamentally different
electromagnetic field tensors. That is, one must separate between 
{\em the active electromagnetic field tensor} which is coupled to gravitation, 
and {\em the passive electromagnetic field tensor} which is not coupled to 
gravity but is relevant for the equations of motion via the Lorentz force law. 
Thirdly, from dimensional analysis, the coupling between charge (and thus the 
active electromagnetic field tensor) and gravitation must in general be 
distinct from the coupling between gravitation and material matter fields. This
means that {\em electromagnetism necessarily couples to space-time geometry 
via a separate coupling parameter}. See refs. [1, 2] for details. These three 
features are characteristic for a formulation of classical electrodynamics in 
quasi-metric space-time.

The main goal of this paper is to formulate classical electrodynamics coupled
to gravity in a way consistent with the QMF. In other words, the question is 
how the general features of the QMF dictate the behaviour of classical 
electrodynamics (i.e., Maxwell's equations) in quasi-metric space-time. To 
answer this question, it is shown in section 3 that it is possible to derive 
the usual Maxwell equations in electrovacuum for the passive electromagnetic 
field in curved space-time from local conservation laws. This motivates the
general definitions of Maxwell's equations within the QMF. Using these 
definitions, the relationship between the passive and active electromagnetic 
field tensors yields an initial-value formulation of electrodynamics coupled 
to gravity. A specific example of this is given in section 4, where exact 
solutions are found for the gravitational and electric fields in the 
electrovacuum outside a spherically symmetric, metrically static charged 
source. Finally, in section 5 it is shown that the main results of geometric 
optics are valid in quasi-metric space-time. That is, it is shown how to derive
the main results of geometric optics (e.g., that light rays are null geodesics)
from Maxwell's equations in curved space-time the way it is done in General 
Relativity (GR).

At the present time, it is not known whether or not quasi-metric gravity is 
viable according to observational criteria (the current observational status 
of the QMF is discussed thoroughly elsewhere [1-3]), and a discussion of that 
is outside the scope of this paper. Besides, just as for GR, fact is that the 
active aspects of the electromagnetic field do not have any consequences which 
are not immeasurably small in practice (but one exception could be the 
dependence on source composition of local gravitational experiments measuring 
the gravitational ``constant''). However, to be viable according to 
theoretical criteria, it is necessary that the QMF accommodates 
electromagnetism in a consistent and natural way; that is what is shown in 
this paper.
\section{Some relevant aspects of the QMF}
Quasi-metric theory and some of its predictions are described in detail
elsewhere [1, 2]; here we merely repeat the basics and the relevant 
formulae. 

The geometrical basis of the QMF consists of a five-dimensional differentiable
manifold with topology ${\cal M}{\times}{\bf R}_1$, where ${\cal M}={\cal S}
{\times}{\bf R}_2$ is a four-dimensional Lorentzian space-time manifold,
${\bf R}_1$ and ${\bf R}_2$ are two copies of the real line and ${\cal S}$ is 
a three-dimensional compact manifold (without boundaries). This geometrical 
structure implies that there exists one extra (degenerate) time dimension
represented by the {\em global time function $t$} as a global coordinate on
${\bf R}_1$ in addition to a ``preferred'' ordinary global time coordinate
$x^0$ on ${\bf R}_2$ with the property that $x^0$ scales like $ct$. That is,
when $t$ is given, so is the ``preferred'' time coordinate $x^0$. Using this
``preferred'' global time coordinate, the four-dimensional quasi-metric 
space-time manifold ${\cal N}$ is constructed by slicing the submanifold 
determined by the equation $x^0=ct$ out of ${\cal M}{\times}{\bf R}_1$. 

Moreover, by construction ${\cal N}$ is equipped with a one-parameter 
family of Lorentzian 4-metrics ${\bf g}_t$. One may alternatively regard 
${\bf g}_t$ as a degenerate five-dimensional metric on a subset of 
${\cal M}{\times}{\bf R}_1$. A special set of coordinate systems especially 
well adapted to the structure of quasi-metric space-time is the set of 
{\em global time coordinate systems (GTCSs)} defined by the condition that 
$x^0$ is a ``preferred'' time coordinate in ${\cal N}$. The set of spatial 
submanifolds ${\cal S}$ taken at the set of constant $t$-values is called 
{\em the fundamental hypersurfaces (FHSs)} and represents a ``preferred'' 
notion of space. Observers always moving orthogonally to the FHSs are called 
{\em fundamental observers (FOs).}

The main physical role of the degenerate dimension represented by $t$ is to
describe global scale changes between gravitational and non-gravitational
systems. The reason one needs a degenerate dimension to describe this, is that 
such global scale changes should not have anything to do with space-time's
causal structure, which is why the evolution of quantities with $t$ is called 
``non-kinematical''. In particular this yields an alternative, non-kinematical
description of the global cosmic expansion.

Now a particular property of the QMF is that the metric family ${\bf g}_t$
does not represent solutions of gravitational field equations. Rather,
${\bf g}_t$ is constructed from a second metric family ${\bf {\bar g}}_t$. The 
transformation ${\bf {\bar g}}_t{\rightarrow}{\bf g}_t$ takes the form of a 
deformation of ${\bf {\bar g}}_t$ along a 3-vector field 
${\bf {\bar x}}_{\cal F}$ [1]. In fact the transformation ${\bf {\bar g}}_t
{\rightarrow}{\bf g}_t$ is just a special case of a more general transformation
of tensor field families. Moreover, this general transformation applies to any 
tensor field which norm is required to be preserved under the transformation 
${\bf {\bar g}}_t{\rightarrow}{\bf g}_t$. Here, we list the transformation 
formulae ${\bf {\bar Z}}_t{\rightarrow}{\bf Z}_t$ valid for a covector family 
${\bf {\bar Z}}_t$. Expressed in a GTCS, we have that (in component notation, 
using Einstein's summation convention where Latin indices take integer values
in the interval 1-3) [1, 2]
\eqa
Z_{(t)0}={\Big (}1-{\frac{v^2}{c^2}}{\Big )}{\bar Z}_{(t)0}, \qquad
Z_{(t)j}={\bar Z}_{(t)j}+{\frac{2{\frac{v}{c}}}{1-{\frac{v}{c}}}}
({\bar e}^s_{\cal F}{\bar Z}_{(t)s}){\bar {\omega}}_{{\cal F}j},
\ena
where ${\bf {\bar e}}_{\cal F}{\equiv}{\frac{t_0}{t}}{\bar e}^s_{\cal F}
{\frac{\partial}{{\partial}x^s}}$ is the unit vector field family (with 
corresponding covector field family ${\bf {\bar {\omega}}}_{\cal F}
{\equiv}{\frac{t}{t_0}}{\bar {\omega}}_{{\cal F}s}dx^s$) along 
${\bf {\bar x}}_{\cal F}$ and $v$ is a scalar field. For the general form of 
$v$, see [1, 2]. For completeness, we also list the transformation formulae 
${\bf {\bar W}}_t{\rightarrow}{\bf W}_t$ valid for a rank 2 tensor family 
${\bf {\bar W}}_t$. These formulae read [1, 2]
\eqa
W_{(t)00}={\Big (}1-{\frac{v^2}{c^2}}{\Big )}^2{\bar W}_{(t)00}, 
\ena
\eqa
W_{(t)0j}={\Big (}1-{\frac{v^2}{c^2}}{\Big )}{\Big [}
{\bar W}_{(t)0j}+{\frac{2{\frac{v}{c}}}{1-{\frac{v}{c}}}}({\bar e}^s_{\cal F}
{\bar W}_{(t)0s}){\bar {\omega}}_{{\cal F}j}{\Big ]},
\ena
\eqa
W_{(t)ij}={\bar W}_{(t)ij}+
{\frac{2{\frac{v}{c}}}{(1-{\frac{v}{c}})^2}}{\bar e}^k_{\cal F}
({\bar {\omega}}_{{\cal F}i}{\bar W}_{(t)kj}
+{\bar W}_{(t)ik}{\bar {\omega}}_{{\cal F}j}).
\ena
The transformation ${\bf {\bar g}}_t{\rightarrow}{\bf g}_t$ may be found
from equations (2)-(4) as a special case.

To make matters as simple as possible, the topology of the FHSs is required to 
be simply connected in addition to being compact. Besides, the geometry of the 
FHSs in $({\cal N},{\bf {\bar g}}_t)$ is postulated to represent a direct 
measure of gravitational length scales as measured in atomic units via a
scale factor family ${\bar F}_t$. To avoid introducing any extra arbitrary 
scale or parameter, ${\bar F}_t$ should be proportional to $ct$. This is the 
global part of ${\bar F}_t$ describing the cosmic expansion, but there is also 
a local part due to gravitation, so we may define 
${\bar F}_t{\equiv}ct{\bar N}_t$, where ${\bar N}_t$ is the lapse function 
field family of the FOs in $({\cal N},{\bf {\bar g}}_t)$. This means that, 
expressed in an arbitrary GTCS, ${\bf {\bar g}}_t$ can be written in a form 
where the explicit $t$-dependence is included via ${\bar F}_t$. This form is 
[1] (using the signature ($-+++$))
\eqa
{\overline {ds}}_t^2={\bar N}_t^2{\Big \{ }
[{\bar N}_{(t)}^k{\bar N}_{(t)}^s{\tilde h}_{(t)ks}-1](dx^0)^2+
2{\frac{t}{t_0}}{\bar N}_{(t)}^k{\tilde h}_{(t)ks}dx^sdx^0+
{\frac{t^2}{t_0^2}}{\tilde h}_{(t)ks}dx^kdx^s{\Big \} },
\ena
where ${\frac{t_0}{t}}{\bar N}_{(t)}^s{\frac{\partial}{{\partial}x^s}}$ is the 
family of shift vectors in the chosen GTCS and where $t_0$ denotes some 
constant reference epoch, usually taken to be the present one. Also 
${\bf {\bar h}}_t$, with components ${\bar h}_{(t)ks}{\equiv}{\frac{t^2}{t_0^2}}
{\bar N}_t^2{\tilde h}_{(t)ks}$, is the metric family intrinsic to the FHSs. A 
complement to equation (5) is the general form for the family ${\bf g}_t$, 
given by the family of line elements (using a GTCS)
\eqa
{ds}_t^2=[N_{(t)}^kN_{(t)}^s{\hat h}_{(t)ks}-N^2](dx^0)^2+
2{\frac{t}{t_0}}N_{(t)}^k{\hat h}_{(t)ks}dx^sdx^0+
{\frac{t^2}{t_0^2}}{\hat h}_{(t)ks}dx^kdx^s,
\ena
where the symbols have similar meanings to their (barred) counterparts in 
equation (5) (the counterpart to ${\bar h}_{(t)ks}$ is
$h_{(t)ks}{\equiv}{\frac{t^2}{t_0^2}}{\hat h}_{(t)ks}$). Note that the lapse
function $N$ is required not to depend on $t$.

Furthermore, the family ${\bf {\bar g}}_t$ is required to be a solution of the 
field equations (in component notation using a GTCS, where the symbol 
'${\mid}$' denotes a space covariant derivative and a comma denotes a partial 
derivative)
\eqa
2{\bar R}_{(t){\bar {\perp}}{\bar {\perp}}}=
{\kappa}^{\rm B}(T^{\rm (EM)}_{(t){\bar {\perp}}{\bar {\perp}}}
+{\hat T}_{(t)s}^{{\rm (EM)}s})+
{\kappa}^{\rm S}(T^{\rm (MA)}_{(t){\bar {\perp}}{\bar {\perp}}}
+{\hat T}_{(t)s}^{{\rm (MA)}s}),
\ena
\eqa
{\bar R}_{(t)j{\bar {\perp}}}+{\Big (}{\frac{{\bar h}_{(t)}^{ik}}{{\bar N}_t}}
{\frac{\partial}{{\partial}x^0}}{\bar h}_{(t)ij}{\Big )}_{{\mid}k}-
{\Big (}{\frac{{\bar h}_{(t)}^{ik}}{{\bar N}_t}}
{\frac{\partial}{{\partial}x^0}}{\bar h}_{(t)ik}{\Big )},_j 
={\kappa}^{\rm B}T^{{\rm (EM)}}_{(t)j{\bar {\perp}}}
+{\kappa}^{\rm S}T^{\rm (MA)}_{(t)j{\bar {\perp}}}.
\ena
Here, ${\bf T}_t$ is the active stress-energy tensor field family (in which 
active mass-energy is treated as a scalar field) split up into one 
electromagnetic part ${\bf T}^{\rm (EM)}_t$ and one part ${\bf T}_t^{\rm (MA)}$
representing material sources. Also, ${\bf {\bar R}}_t$ is the Ricci tensor 
field family calculated from ${\bf {\bar g}}_t$. Moreover,
${\kappa}^{\rm B}{\equiv}{\frac{8{\pi}G^{\rm B}}{c^4}}$ and
${\kappa}^{\rm S}{\equiv}{\frac{8{\pi}G^{\rm S}}{c^4}}$, where the ``bare'' 
gravitational ``constant'' $G^{\rm B}$ couples to ${\bf T}_t^{\rm (EM)}$ and the 
``screened'' gravitational ``constant'' $G^{\rm S}$ couples to 
${\bf T}_t^{\rm (MA)}$. These ``constants'' are by convention set to the values 
they would have had in an empty Universe at the arbitrary epoch $t_0$
(as measured in local gravitational experiments). Finally, the symbol 
`${\bar {\perp}}$' denotes a projection with the negative normal unit vector 
field family $-{\bf {\bar n}}_t$ of the FHSs. (A ``hat'' above an object 
denotes an object projected into the FHSs.) 

The field equations are completed by requiring that the traceless quantity 
${\bar Q}_{(t)ij}$ defined from the relationship [1, 2]
\eqa
{\bar G}_{(t)ij}=-{\bar Q}_{(t)ij}-2c^{-2}{\bar a}_{{\cal F}i{\mid}j}-
2c^{-4}{\bar a}_{{\cal F}i}{\bar a}_{{\cal F}j}
-2{\bar K}_{(t)i}^{s}{\bar K}_{(t)sj}+2{\bar K}_t{\bar K}_{(t)ij}
\nonumber \\
+{\frac{1}{3}}{\Big [}2{\bar R}_{(t){\bar {\perp}}{\bar {\perp}}}
-{\bar G}_{(t){\bar {\perp}}{\bar {\perp}}}+2c^{-2}{\bar a}_{{\cal F}{\mid}s}^s
+2c^{-4}{\bar a}_{{\cal F}}^s{\bar a}_{{\cal F}s}
+2{\bar K}_{(t)ks}{\bar K}_{(t)}^{ks}-2{\bar K}_t^2
{\Big ]}{\bar h}_{(t)ij},
\ena
should vanish. Here, ${\bf {\bar a}}_{\cal F}$ is the 4-acceleration field of 
the FOs and ${\bf {\bar G}}_t$ is the Einstein tensor family. Moreover, 
${\bf {\bar K}}_t$ (with trace ${\bar K}_t$) is the extrinsic tensor family of 
the FHSs in $({\cal N},{\bf {\bar g}}_t)$. Equation (9), in addition to general
expressions [1, 2] valid for the projections 
${\bar R}_{(t){\bar {\perp}}{\bar {\perp}}}$, ${\bar G}_{(t){\bar {\perp}}{\bar {\perp}}}$ 
and ${\bar G}_{(t)ij}$, then yield the remainder quasi-metric field equation
\eqa
{\bar Q}_{(t)ij}{\equiv}{\frac{1}{{\bar N}_t}}{\cal L}_{{\bar N}_t
{\bf {\bar n}}_t}{\bar K}_{(t)ij}+
{\frac{1}{3}}{\Big [}2{\bar K}_{(t)ks}{\bar K}_{(t)}^{ks}-{\bar K}_t^2
-{\cal L}_{{\bf {\bar n}}_t}{\bar K}_t{\Big ]}{\bar h}_{(t)ij}
+{\bar K}_t{\bar K}_{(t)ij}
\nonumber \\
-c^{-2}{\bar a}_{{\cal F}i{\mid}j}-
c^{-4}{\bar a}_{{\cal F}i}{\bar a}_{{\cal F}j}
+{\Big [}c^{-2}{\bar a}_{{\cal F}{\mid}s}^s
-{\frac{1}{(ct{\bar N}_t)^2}}{\Big ]}{\bar h}_{(t)ij}-{\bar H}_{(t)ij}=0,
\ena
where the prior-geometric requirement on the spatial Ricci curvature scalar 
family ${\bar P}_t$,
\eqa
{\bar P}_t=-4c^{-2}{\bar a}_{{\cal F}{\mid}s}^s
+2c^{-4}{\bar a}_{{\cal F}}^s{\bar a}_{{\cal F}s}+{\frac{6}{(ct{\bar N}_t)^2}},
\ena
ensures that equation (10) is indeed manifestly traceless. In equation (10), 
${\bf {\bar H}}_t$ is the spatial Einstein tensor family intrinsic to the FHSs 
and ${\cal L}_{{\bf {\bar n}}_t}$ denotes a projected Lie derivative in the 
direction normal to the FHSs. We notice that equation (11) eliminates any 
possible gauge freedom in determining lapse and shift for the FOs, ensuring 
that the time evolution of the FHSs cannot be ambiguous (given suitable initial 
conditions). We also notice that equation (9) is equivalent to the criterion 
[1, 2]
\eqa
{\bar C}_{(t){\bar {\perp}}i{\bar {\perp}}j}={\tilde H}_{(t)ij}+
{\frac{1}{(ct{\bar N}_t)^2}}{\bar h}_{(t)ij},
\ena
involving a particular projection of the Weyl tensor family ${\bf {\bar C}}_t$
and the spatial Einstein tensor family ${\bf {\tilde H}}_t$ obtained from the
spatial metric family ${\bf {\tilde h}}_t$, where
${\bf {\tilde h}}_t{\equiv}{\frac{t^2_0}{t^2}}{\bar N}_t^{-2}{\bf {\bar h}}_t$.

Next, local conservation laws for ${\bf T}_t$ in $({\cal N},{\bf {\bar g}}_t)$
read (using Einstein's summation convention where Greek indices take integer 
values in the interval 0-3)
\eqa
T_{(t){\mu};{\alpha}}^{\alpha}=2{\frac{{\bar N}_t,_{\nu}}{{\bar N}_t}}
T_{(t){\mu}}^{\nu}, \qquad T_{(t){\mu}{\bar *}t}^0=
-{\frac{2}{{\bar N}_t}}{\Big (}{\frac{1}{t}}+{\frac{{\bar N}_t,_t}{{\bar N}_t}}
{\Big )}T_{(t){\bar{\perp}}{\mu}},
\ena
where the symbol `${\bar *}$' denotes a covariant derivative found from the 
five-dimensional connection ${\topstar{\bf {\bar {\nabla}}}}$ compatible with 
${\bf {\bar g}}_t$, and a semicolon denotes a metric covariant derivative found 
from the family of Levi-Civita connections ${\bf {\bar {\nabla}}}$ compatible 
with single members of the family ${\bf {\bar g}}_t$. Note that, unlike its 
counterpart in GR, ${\bf {\bar {\nabla}}}{\cdot}{\bf T}_t$ does not vanish in 
general. Finally, the equations of motion in $({\cal N},{\bf g}_t)$ take the 
form (in a GTCS)
\eqa
{\frac{d^2x^{\alpha}}{d{\lambda}^2}}+{\Big (}
{\topstar{\Gamma}}^{\alpha}_{t{\beta}}{\frac{dt}{d{\lambda}}}+
{\topstar{\Gamma}}^{\alpha}_{{\sigma}{\beta}}{\frac{dx^{\sigma}}{d{\lambda}}}
{\Big )}{\frac{dx^{\beta}}{d{\lambda}}}={\Big (}{\frac{d{\tau}_t}{d{\lambda}}}
{\Big )}^2a_{(t)}^{\alpha},
\ena
\eqa
{\topstar{\Gamma}}^{\alpha}_{t{\beta}}{\equiv}
{\topstar{\Gamma}}^{\alpha}_{{\beta}t}{\equiv}
{\frac{1}{t}}{\delta}^{\alpha}_s{\delta}_{\beta}^s+
{\frac{1}{2}}{\delta}^{\alpha}_i{\delta}_{\beta}^k
{\hat h}^{is}_{(t)}{\hat h}_{(t)ks},_t, \qquad
{\topstar{\Gamma}}^{\alpha}_{{\sigma}{\beta}}{\equiv}{\frac{1}{2}}
g_{(t)}^{{\alpha}{\rho}}{\Big (}g_{(t){\rho}{\beta},{\sigma}}+
g_{(t){\sigma}{\rho},{\beta}}-g_{(t){\sigma}{\beta},{\rho}}{\Big )},
\ena
where ${\tau}_t$ is the proper time measured along a time-like curve and
${\lambda}$ is a general affine parameter. (Here, the $a_{(t)}^{\alpha}$ are the
components of the family of 4-accelerations ${\bf a}_t$.)
\section{Active and passive electromagnetic field tensors}
Within the QMF, the active representation of some non-gravitational 
field (i.e., the representation describing its coupling to gravity), will not
be equal to the passive representation. That is, the form non-gravitational
fields take in non-gravitational force laws, will not be identical to the form 
non-gravitational fields take when contributing to the active stress-energy 
tensor. For electromagnetism in particular, we will illustrate below how these 
representations must be different to be consistent with the equations of 
motion in $({\cal N},{\bf {\bar g}}_t)$ (and in $({\cal N},{\bf g}_t)$).

We proceed to define two different representations of the electromagnetic field
in quasi-metric space-time. First, we define {\em the passive electromagnetic
field tensor family ${\bf {\bar F}}_t$} in $({\cal N},{\bf {\bar g}}_t)$ via a
vector potential family ${\bf {\bar A}}_t$ in $({\cal N},{\bf {\bar g}}_t)$.
That is, for reasons given below, we define ${\bf {\bar F}}_t$ via its 
components expressed in a GTCS
\eqa
{\bar F}_{(t)0j}&=&{\frac{t_0}{t}}{\bar A}_{(t)j;0}-{\bar A}_{(t)0;j}=
{\frac{t_0}{t}}{\bar A}_{(t)j,0}-{\bar A}_{(t)0,j}=-{\bar F}_{(t)j0}, 
\nonumber \\
{\bar F}_{(t)ij}&=&{\bar A}_{(t)j;i}-
{\bar A}_{(t)i;j}={\bar A}_{(t)j,i}-{\bar A}_{(t)i,j}=-{\bar F}_{(t)ji}.
\ena
Now we define the $t$-dependence of ${\bf {\bar A}}_t$ from the requirement
${\bf {\stackrel{\star}{\bar {\nabla}}}}_{\frac{\partial}{{\partial}t}}
{\bf {\bar A}}_t=-({\frac{1}{t}}+{\frac{{\bar N}_{t,t}}{{\bar N}_t}})
{\bf {\bar A}}_t$, so that
\eqa
{\bar A}_{(t)0},_t=-{\frac{1}{t}}{\bar A}_{(t)0}, \quad
{\bar A}_{(t)j},_t={\frac{1}{2}}{\bar A}_{(t)s}{\tilde h}_{(t)}^{sk}
{\tilde h}_{(t)kj},_t,
\quad 
{\mid}{\bf {\bar A}}_t{\mid},_t= 
-({\frac{1}{t}}+{\frac{{\bar N}_{t,t}}{{\bar N}_t}})
{\mid}{\bf {\bar A}}_t{\mid}.
\ena 
Here, ${\mid}{\bf {\bar A}}_t{\mid}{\equiv}{\sqrt{{\mid}{\bar A}_{(t){\mu}}
{\bar A}_{(t)}^{\mu}{\mid}}}$ denotes the norm of ${\bf {\bar A}}_t$. The 
reason for the specific dependences on $t$ of ${\bf {\bar A}}_t$ and 
${\bf {\bar F}}_t$ is that within the QMF, the passive electromagnetic field 
by construction experiences a cosmic attenuation due to the global cosmic 
expansion. This cosmic attenuation represents global scale changes between 
electromagnetism and gravitation. Moreover, these global scale changes are 
physically realized via the explicit $t$-dependence of the norm of 
${\bf {\bar F}}_t$; we see from equations (16) and (17) that its norm 
attenuates according to the formula ${\mid}{\bf {\bar F}}_t{\mid},_t{\equiv}
{\sqrt{{\mid}{\bar F}_{(t){\sigma}{\rho}}{\bar F}_{(t)}^{{\sigma}{\rho}}
{\mid}}},_t=-2({\frac{1}{t}}+{\frac{{\bar N}_{t,t}}{{\bar N}_t}}){\mid}
{\bf {\bar F}}_t{\mid}+f({\bf {\tilde h}}_t,_t)$,
where $f$ is a complicated function irrelevant to said global cosmic 
attenuation.

To predict the passive aspects of electromagnetism according to the QMF, we 
must also construct the ``physical'' passive electromagnetic field tensor 
family ${\bf F}_t$ in $({\cal N},{\bf g}_t)$ from ${\bf {\bar F}}_t$ in exactly
the same way as ${\bf g}_t$ is constructed from ${\bf {\bar g}}_t$ using 
equations (2)-(4). The construction of ${\bf F}_t$ is necessary since it is 
${\bf F}_t$ which enters into the Lorentz force law 
\eqa
ma_{(t)}^{\alpha}={\frac{q}{c}}F_{(t){\nu}}^{\alpha}u_{(t)}^{\nu},
\ena
in $({\cal N},{\bf g}_t)$, where ${\bf u}_t$ is the 4-velocity field of the 
charged matter and where $q$ is the passive charge and $m$ the passive mass of 
a test particle.

Second, we define {\em the active electromagnetic field tensor family 
${\bf {\tilde F}}_t$} in $({\cal N},{\bf {\bar g}}_t)$ directly from
${\bf {\bar F}}_t$:
\eqa
{\bf {\tilde F}}_t{\equiv}{\frac{t}{t_0}}{\bar N}_t{\bf {\bar F}}_t, \qquad
{\tilde F}_{(t){\alpha}{\beta}}{\equiv}{\frac{t}{t_0}}
{\bar N}_t{\bar F}_{(t){\alpha}{\beta}}.
\ena
The $t$-dependence of ${\bf {\tilde F}}_t$ differs from that of 
${\bf {\bar F}}_t$ since the active electromagnetic field will experience a 
global cosmic increase of active mass-energy which will partly cancel the 
effects of the global cosmic attenuation (i.e.,
${\mid}{\bf {\tilde F}}_t{\mid},_t=-({\frac{1}{t}}+{\frac{{\bar N}_{t,t}}
{{\bar N}_t}}){\mid}{\bf {\tilde F}}_t{\mid}+$ irrelevant terms).

The point now is that ${\bf {\tilde F}}_t$ is coupled dynamically to gravity
whereas ${\bf {\bar F}}_t$ is not. That is, ${\bf {\tilde F}}_t$ determines 
the active electromagnetic field stress-energy tensor ${\bf T}_t^{{\rm (EM)}}$ 
via the familiar formula (in component notation)
\eqa
T_{(t){\alpha}{\beta}}^{{\rm (EM)}}={\frac{1}{4{\pi}}}{\Big (}
{\tilde F}_{(t){\alpha}}^{{\ }{\ }{\ }{\,}{\,}{\nu}}
{\tilde F}_{(t){\beta}{\nu}}-{\frac{1}{4}}{\tilde F}_{(t){\sigma}{\rho}}
{\tilde F}_{(t)}^{{\sigma}{\rho}}{\bar g}_{(t){\alpha}{\beta}}{\Big )}.
\ena
Similarly, ${\bf {\bar F}}_t$ defines the passive electromagnetic field 
stress-energy tensor 
${\bf {\bar {\cal T}}}_t^{{\rm (EM)}}$ in $({\cal N},{\bf {\bar g}}_t)$
\eqa
{\bar {\cal T}}_{(t){\alpha}{\beta}}^{{\rm (EM)}}={\frac{1}{4{\pi}}}{\Big (}
{\bar F}_{(t){\alpha}}^{{\ }{\ }{\ }{\,}{\,}{\nu}}
{\bar F}_{(t){\beta}{\nu}}-{\frac{1}{4}}{\bar F}_{(t){\sigma}{\rho}}
{\bar F}_{(t)}^{{\sigma}{\rho}}{\bar g}_{(t){\alpha}{\beta}}{\Big )}.
\ena
To find electromagnetic field equations for ${\bf {\bar F}}_t$, we first 
consider the special case of an electrovacuum. In general, the relationship 
between the active electromagnetic stress-energy tensor ${\bf T}^{\rm (EM)}_t$ 
given in equation (20) and the corresponding passive electromagnetic 
stress-energy tensor ${\bf {\bar {\cal T}}}^{\rm (EM)}_t$ defined in equation 
(21), is given by
\eqa
{\bf T}^{\rm (EM)}_t={\frac{t^2}{t_0^2}}{\bar N}_t^2
{\bf {\bar {\cal T}}}^{\rm (EM)}_t, \qquad \Rightarrow \qquad  
{\bf {\bar {\nabla}}}{\cdot}{\bf {\bar {\cal T}}}_t^{\rm (EM)}=0, \qquad
{\rm (electrovacuum)}
\ena
where the implication follows from equation (13) for electrovacuum. Note 
that the relationship (22) takes the same form as for a perfect fluid of null 
particles [2]. Using equations (21) and (22), we may now perform a standard 
derivation of the usual electrovacuum Maxwell equations 
${\bar F}_{(t){\mu};{\nu}}^{\nu}=0$ in $({\cal N},{\bf {\bar g}}_t)$. Just as
for GR, this derivation acts as a motivation for the general definition of the 
usual Maxwell equations in $({\cal N},{\bf {\bar g}}_t)$, namely
\eqa
{\bar F}_{(t){\alpha}{\beta};{\mu}}+{\bar F}_{(t){\mu}{\alpha};{\beta}}+
{\bar F}_{(t){\beta}{\mu};{\alpha}}={\bar F}_{(t){\alpha}{\beta},{\mu}}+
{\bar F}_{(t){\mu}{\alpha},{\beta}}+{\bar F}_{(t){\beta}{\mu},{\alpha}}=0,
\ena
\eqa
{\bar F}_{(t){\mu};{\nu}}^{\nu}=-{\frac{4{\pi}}{c}}{\bar J}_{(t){\mu}}, \qquad 
{\bar J}_{(t){\mu}}{\equiv}{\varrho}_{\text c}{\bar u}_{(t){\mu}}, \qquad 
\Rightarrow \qquad {\bar J}^{\mu}_{(t);{\mu}}={\frac{1}{\sqrt{-{\bar g}_t}}}
({\sqrt{-{\bar g}_t}}{\bar J}_{(t)}^{\mu}),_{\mu}=0,
\ena
where ${\bf {\bar J}}_t$ is the 4-current of passive charge in $({\cal N},
{\bf {\bar g}}_t)$, ${\bar g}_t$ is the determinant of the metric family
${\bf {\bar g}}_t$ and ${\varrho}_{\text c}$ is the density of passive charge $q$
as measured in the local inertial rest frame of the fluid. Note that only the
metric part of the connection enters into equation (24). This means that
equation (24) has the same properties as its counterpart in metric gravity. In 
particular, it means that equation (24) ensures that passive charge is 
conserved in $({\cal N},{\bf {\bar g}}_t)$ and that we in general have
\eqa
{\bf {\bar {\cal T}}}_{(t){\alpha}{\,}{\,}{\,}{\,}{\,}{\,}{\,}{\,}
{\,};{\beta}}^{{\rm (EM)}{\beta}}=
-{\frac{1}{c}}{\bar F}_{(t){\alpha}{\beta}}{\bar J}_{(t)}^{\beta}, \qquad
{\rm (general{\,}{\,}case)}
\ena
rather than the special case shown in equation (22). Now one important point 
is that no counterparts to equations (23), (24) exist in 
$({\cal N},{\bf g}_t)$. However, it is still possible to define the 4-current 
of passive charge ${\bf J}_t$ in $({\cal N},{\bf g}_t)$:
\eqa
{\bf J}_t{\equiv}
{\sqrt{{\frac{{\bar h}_t}{h_t}}}}{\varrho}_{\rm c}{\bf u}_t, \qquad
\Rightarrow \qquad J_{(t){\perp}}{\sqrt{h_t}}={\bar J}_{(t){\bar {\perp}}}
{\sqrt{{\bar h}_t}},
\ena
where $J_{(t){\perp}}{\equiv}-n_{(t){\mu}}J_{(t)}^{\mu}$ (and ${\bf n}_t$ is 
the unit normal vector field family to the FHSs in $({\cal N},{\bf g}_t)$). 
Moreover, ${\bar h}_t$ and $h_t$ are the determinants of the spatial metric 
families ${\bf {\bar h}}_t$ and ${\bf h}_t$, respectively. The square root 
factor is necessary in the definition of ${\bf J}_t$ to ensure that the amount
of passive charge is identical in $({\cal N},{\bf {\bar g}}_t)$ and in 
$({\cal N},{\bf g}_t)$. The implication in equation (26) follows because the 
norm of ${\bf {\bar u}}_t$ (and of ${\bf {\bar n}}_t$) must be invariant under
the transformation ${\bf {\bar g}}_t{\rightarrow}{\bf g}_t$. This means that 
${\bf {\bar u}}_t{\rightarrow}{\bf u}_t$ and 
${\bf {\bar n}}_t{\rightarrow}{\bf n}_t$, i.e., that ${\bf {\bar u}}_t$ and 
${\bf {\bar n}}_t$ transform according to equation (1). Using Gauss' theorem 
within the 4-dimensional coordinate volume bounded by two different FHSs and 
chosen such that ${\bf {\bar J}}_t$ vanishes on lateral boundaries, it is now 
straightforward to show that equation (24) ensures that passive charge is 
conserved for said volume in $({\cal N},{\bf {\bar g}}_t)$. But then equation 
(26) implies that passive charge is conserved for said volume in 
$({\cal N},{\bf g}_t)$ as well.

For weak electromagnetic fields, it is a convenient approximation to neglect 
the effects of electromagnetism on space-time geometry. This is equivalent to
considering passive electromagnetism only in an independent quasi-metric curved
space-time background. For this case, initial-value calculations may be done as 
follows: given ${\varrho}_{\rm c}$ and ${\bf {\bar u}}_{t_1}$ (found from 
${\bf u}_{t_1}$) for some initial value $t=t_1$, it is possible to calculate 
${\bf {\bar F}}_{t_1}$ from equations (23) and (24). Then ${\bf F}_{t_1}$ can 
be found from ${\bf {\bar F}}_{t_1}$ just as ${\bf g}_{t_1}$ can be found from 
${\bf {\bar g}}_{t_1}$, using equations (2)-(4). Moreover, the motion of the 
source charges can be found from the geodesic equation (14) in 
$({\cal N},{\bf g}_t)$ (using equation (18)). This again yields 
${\varrho}_{\rm c}$ and ${\bf u}_{(t_1+{\Delta t})}$ (from which 
${\bf {\bar u}}_{(t_1+{\Delta}t)}$ can be calculated) for some later time 
$t=t_1+{\Delta}t$. The procedure can then be repeated to find the evolution of 
the electromagnetic field at progressively later times. We thus have a 
well-defined initial value problem for passive electromagnetism in quasi-metric
space-time. Note that the transformations (2)-(4) preserve norm by 
construction, so if, e.g., ${\bf {\bar F}}_t$ is null, so is ${\bf F}_t$. This 
means that passive electrodynamics in quasi-metric space-time respects the 
local light cone.

However, in general one must also take into account the active aspects of
electromagnetism. To do that, we may we use the definition (19) together with 
equation (24) to find field equations for the active electromagnetic field:
\eqa
{\tilde F}_{(t){\mu};{\nu}}^{\nu}=-{\frac{4{\pi}}{c}}{\tilde J}_{(t){\mu}}+
{\frac{{\bar N}_t,_{\nu}}{{\bar N}_t}}{\tilde F}_{(t){\mu}}^{\nu}, \qquad 
{\tilde J}_{(t){\mu}}{\equiv}{\tilde {\varrho}}_{\text c}{\bar u}_{(t){\mu}}, 
\qquad \Rightarrow \qquad {\tilde J}^{\mu}_{(t);{\mu}}=
{\frac{{\bar N}_t,_{\nu}}{{\bar N}_t}}{\tilde J}_{(t)}^{\nu}.
\ena
Here, the 4-current of active charge ${\bf {\tilde J}}_t$ in
$({\cal N},{\bf {\bar g}}_t)$ is defined by
\eqa
{\bf {\tilde J}}_t{\equiv}{\tilde{\varrho}}_{\text c}{\bf {\bar u}}_t=
{\frac{t}{t_0}}{\bar N}_t{\varrho}_{\text c}{\bf {\bar u}}_t,
\ena
where the last step follows from equation (29) below, and where
${\tilde{\varrho}}_{\text c}$ is the density of active charge $q_t$. Analogous
to active mass, active charge is a scalar field and it describes how charge 
couples to gravity. How active charge varies in quasi-metric space-time is 
found from dimensional analysis [2]. That is, just as one may infer the 
variability of the active mass $m_t$ from the gravitational quantity 
$G^{\rm S}m_t/c^2$ (which has the dimension of length), one may infer the 
variability of the active charge from the gravitational quantity 
$G^{\rm B}q_t^2/c^4$ which has the dimension of length squared. The result is
\eqa
q_t,_{\mu}= {\frac{{\bar N}_t,_{\mu}}{{\bar N}_t}}q_t, \qquad
q_t,_t={\Big (}{\frac{1}{t}}+{\frac{{\bar N}_t,_t}{{\bar N}_t}}{\Big )}q_t, 
\qquad \Rightarrow \qquad q_t={\frac{t}{t_0}}{\bar N}_tq.
\ena
We see from equation (27) that active charge is not necessarily conserved, nor 
is there any reason that it should be. 

Finally, we notice that the exterior derivative of ${\bf {\tilde F}}_t$ will 
not vanish in general, so there is no counterpart to equation (23) valid for
${\bf {\tilde F}}_t$. Also note that equations (19), (20), (21), (25) and 
(28) yield that
\eqa
T_{(t){\alpha}{\,}{\,}{\,}{\,}{\,}{\,}{\,}{\,}{\,};{\beta}}^{{\rm (EM)}{\beta}}
=2{\frac{{\bar N}_{t,{\beta}}}{{\bar N}_t}}T_{(t){\alpha}}^{{\rm (EM)}{\beta}}
-{\frac{1}{c}}{\tilde F}_{(t){\alpha}{\beta}}{\tilde J}_{(t)}^{\beta}.
\ena
When the active stress-energy tensor of the charged matter source is added to 
equation (30), we get back equation (13) for the total active stress-energy 
tensor.
 
The description of how electromagnetism couples to gravity in quasi-metric 
space-time is now given from the definition (19) and the usual Maxwell 
equations (23) and (24) in curved space-time. These equations are coupled to 
the gravitational field equations (7), (8) and (10) and should be solved 
simultaneously. Given suitable initial conditions, it is thus the coupled 
system of equations (7), (8), (10), (14), (18), (23) and (24) which must be 
solved to find the fields ${\bf {\bar F}}_t$ and ${\bf {\bar g}}_t$ for each 
time step. Besides, to find these fields at progressively later times, it is 
also necessary to calculate ${\bf g}_t$ and ${\bf F}_t$ for each time step by 
the method described in [1, 2] using equations (2)-(4). It is thus possible to 
set up a well-defined initial value problem for electromagnetism coupled to 
gravitation in quasi-metric space-time. Note that active electromagnetism in 
quasi-metric space-time is also guaranteed to respect the local light cone 
since ${\bf {\tilde F}}_t$ is null if and only if ${\bf {\bar F}}_t$ is null 
(this follows from equation (19)).
\section{The spherically symmetric, metrically static case}
\subsection{Electrovacuum outside a charged source}
In this section, we set up the equations valid for the gravitational and
electromagnetic fields outside a spherically symmetric, charged source. It is
required that the system is ``metrically static'', i.e., that the only time
dependence is via the effect on the spatial geometry of the global cosmic
expansion. We then find exact solutions of these equations. (It is also 
required that the system is at rest with respect to the cosmic rest frame, but 
this requirement is not crucial, see references [1-3] for a discussion of this 
point.)

To begin with, we notice that the extrinsic curvature tensor family 
${\bf {\bar K}}_t$ vanishes identically for metrically static systems [1, 2], 
so we find from equations (10) and (11) that equation (5) takes a special form 
for these cases. That is, we introduce a spherical GTCS 
${\{}x^0,{\rho},{\theta},{\phi}{\}}$ which with respect to the FOs are at 
rest and where ${\rho}$ is the radial coordinate. Then a metrically static, 
spherically symmetric metric family ${\bf {\bar g}}_t$ may, without loss of 
generality, be expressed in the form (as a special case of equation (5))
\eqa
{\overline {ds}}_t^2={\bar B}({\rho}){\Big [}-(dx^0)^2+
{\frac{t^2}{t_0^2}}{\Big (}{\frac{d{\rho}^2}{1-{\frac{{\rho}^2}{{\Xi}_0^2}}}}
+{\rho}^2d{\Omega}^2{\Big )}{\Big ]},
\ena
where ${\bar B}{\equiv}{\bar N}_t^2$, $d{\Omega}^2{\equiv}d{\theta}^2+
{\sin}^2{\theta}d{\phi}^2$, ${\Xi}_0{\equiv}ct_0$ and $t_0$ is some arbitrary 
reference epoch. The unknown function ${\bar B}({\rho})$ shall be determined 
by solving the field equation (7).

Next we proceed by finding an expression for ${\bf {\tilde F}}_t$. Due to 
symmetry, its only components in the above defined GTCS are given in terms of
the active electric field ${\bf {\tilde E}}_t$ as seen by the
FOs. That is, ${\bf {\tilde E}}_t$ is defined by its radial component
\eqa
{\tilde E}_{(t){\rho}}{\equiv}
{\tilde F}_{(t){\bar {\perp}}{\rho}}{\equiv}-
{\bar n}_{(t)}^{\mu}{\tilde F}_{(t){\mu}{\rho}}
=-{\frac{1}{{\bar N}_t}}{\tilde F}_{(t)0{\rho}}={\tilde E}_{(t_0){\rho}}.
\ena
Since the components of the active magnetic field vanish in the chosen GTCS,
equation (27) yields
\eqa
{\Big (}{\bar N}_t^{-1}{\tilde F}_{(t)}^{0{\rho}}{\Big )}_{;{\rho}}
={\frac{4{\pi}}{c}}{\bar N}_t^{-1}{\tilde J}_{(t)}^0,
\ena
or equivalently,
\eqa
{\frac{\partial}{{\partial}{\rho}}}{\Big (}{\bar N}_t^{-1}{\sqrt{-{\bar g}_t}}
{\tilde F}_{(t)}^{0{\rho}}{\Big )}=4{\pi}{\frac{t}{t_0}}{\sqrt{-{\bar g}_t}}
{\bar N}_t^{-1}{\varrho}_{\text c}.
\ena
We now integrate equation (34) from the origin to some radial coordinate 
${\rho}>{\rho}_{\rm sf}$, where ${\rho}_{\rm sf}$ is the coordinate radius of the 
source. Assuming that all fields are continuous and well-behaved, we get
\eqa
{\rho}^2{\sqrt{1-{\frac{{\rho}^2}{{\Xi}_0^2}}}}{\tilde E}_{(t_0){\rho}}=
4{\pi}{\frac{t^3}{t_0^3}}{\int}_{{\!}{\!}{\!}0}^{{\rho}_{\rm sf}}
{\frac{{\varrho}_{\rm c}{\bar B}^{3/2}{\rho}^2d{\rho}}
{{\sqrt{1-{\frac{{\rho}^2}{{\Xi}_0^2}}}}}}=
{\int}{\int}{\int}{\varrho}_{\text c}{\sqrt{{\bar h}_t}}d^3x{\equiv}Q,
\ena
where the triple integration is taken over the volume of the source. We see
that, since we have integrated the passive charge density over the source 
volume, it is natural to interpret $Q$ as the passive charge of the source, 
which is obviously a constant (since the left hand side of equation (35) has 
no time dependence). From equations (32) and (35) we thus have that
\eqa
{\tilde E}_{(t){\rho}}={\frac{Q}{{\rho}^2
{\sqrt{1-{\frac{{\rho}^2}{{\Xi}_0^2}}}}}}.
\ena
Next we must find expressions for projections of ${\bf T}^{\rm (EM)}_t$ to be put 
into the field equations. Using equations (20), (32) and (36) and doing the 
projections, we find
\eqa
T_{(t){\bar {\perp}}{\bar {\perp}}}^{\rm (EM)}={\frac{1}{8{\pi}}}
{\hat{\tilde F}}_{(t){\bar {\perp}}}^{\rho}{\tilde F}_{(t){\rho}{\bar {\perp}}}
={\frac{1}{8{\pi}}}{\tilde E}_{(t){\rho}}{\hat{\tilde E}}_{(t)}^{\rho}=
{\frac{t_0^2}{t^2}}{\frac{Q^2}{8{\pi}{\bar B}{\rho}^4}}{\equiv}
{\frac{t^2_0}{t^2}}{\bar B}^{-1}{\bar{\varrho}}_{\text m}^{\rm (EM)}c^2,
\ena
\eqa
T_{(t){\rho}}^{{\rm (EM)}{\rho}}=
-T_{(t){\bar {\perp}}{\bar {\perp}}}^{\rm (EM)}{\equiv}
-{\frac{t^2_0}{t^2}}{\bar B}^{-1}{\bar{p}}^{\rm (EM)}=
-{\frac{t^2_0}{t^2}}{\bar B}^{-1}{\bar{\varrho}}_{\text m}^{\rm (EM)}c^2, 
\ena
\eqa
T_{(t){\theta}}^{{\rm (EM)}{\theta}}=T_{(t){\phi}}^{{\rm (EM)}{\phi}}=
-T_{(t){\rho}}^{{\rm (EM)}{\rho}}, \qquad
T_{(t){\bar {\perp}}{\rho}}^{\rm (EM)}=T_{(t){\bar {\perp}}{\theta}}^{\rm (EM)}
=T_{(t){\bar {\perp}}{\phi}}^{\rm (EM)}=0.
\ena
Here, ${\bar{\varrho}}_{\text m}^{\rm (EM)}$ and ${\bar p}^{\rm (EM)}$ are, 
respectively, the so-called ``properly scaled density'' and the corresponding 
``properly scaled pressure'' of active electromagnetic mass-energy. The point 
of introducing these quantities, is that the formal dependence of 
${\bf T}_t^{\rm (EM)}$ on the scale factor ${\bar F}_t$ has been factored out 
(there is such a dependence since ${\bf T}_t^{\rm (EM)}$ has the formal 
dimension of inverse length squared). 

We may now insert equations (37), (38) and (39) into the field equation (7). 
The result is similar to the case of a spherically symmetric, metrically 
static perfect fluid [3], the only difference being that the pressure is not 
isotropic for the electromagnetic field. Using the equation derived in [3], we 
find (with $r_{\text Q0}{\equiv}{\sqrt{2G^{\rm B}}}{\mid}Q{\mid}/c^2$)
\eqa
(1-{\frac{{\rho}^2}{{\Xi}_0^2}}){\frac{{\bar B},_{{\rho}{\rho}}}{\bar B}}+
{\frac{2}{\rho}}(1-{\frac{3{\rho}^2}{2{\Xi}_0^2}})
{\frac{{\bar B},_{\rho}}{\bar B}}={\frac{r_{\rm Q0}^2}{{\rho}^4}}.
\ena
An (unique) exact solution of equation (40), given the correspondence with the
uncharged case [3], is
\eqa
{\bar B}({\rho})={\cosh}{\Big [}{\frac{r_{\rm Q0}}{\rho}}{\sqrt{1-
{\frac{{\rho}^2}{{\Xi}_0^2}}}}{\Big ]}-{\frac{r_{\rm s0}}{r_{\rm Q0}}}
{\sinh}{\Big [}{\frac{r_{\rm Q0}}{\rho}}
{\sqrt{1-{\frac{{\rho}^2}{{\Xi}_0^2}}}}{\Big ]}, \qquad
{\rho}_{\rm sf}{\leq}{\rho}<{\Xi}_0,
\ena
where the constant $r_{\rm s0}$ is found by integrating equation (7) once and 
then comparing with ${\bar B},_{\rho}$ obtained from equation (41). The result 
is
\eqa
r_{\text s0}{\equiv}{\Big (}{\frac{2M^{\rm (MA)}_{t_0}G^{\rm S}}{c^2}}+
{\frac{2M^{\rm (EM)}_{t_0}G^{\rm B}}{c^2}}{\Big )}{\rm sech}
{\Big [}{\frac{r_{\rm Q0}}{{\rho}_{\rm sf}}}
{\sqrt{1-{\frac{{\rho}_{\rm sf}^2}{{\Xi}_0^2}}}}{\Big ]}+
r_{\rm Q0}{\tanh}{\Big [}{\frac{r_{\rm Q0}}
{{\rho}_{\rm sf}}}{\sqrt{1-{\frac{{\rho}_{\rm sf}^2}{{\Xi}_0^2}}}}{\Big ]},
\ena
\eqa
M^{\rm (MA)}_{t_0}{\equiv}c^{-2}{\int}{\int}{\int}{\bar N}_{t_0}{\Big [}
T^{\rm (MA)}_{(t_0){\bar {\perp}}{\bar {\perp}}}+
{\hat T}^{{\rm (MA)}s}_{(t_0)s}{\Big ]}d{\bar V}_{t_0}, \nonumber \\
M^{\rm (EM)}_{t_0}{\equiv}c^{-2}{\int}{\int}{\int}{\bar N}_{t_0}{\Big [}
T^{\rm (EM)}_{(t_0){\bar {\perp}}{\bar {\perp}}}+
{\hat T}^{{\rm (EM)}s}_{(t_0)s}{\Big ]}d{\bar V}_{t_0}.
\ena
Here, the integration is taken over the central source (i.e., 
${\rho}{\leq}{\rho}_{\rm sf}$) and ${\bf T}_t^{\rm (MA)}+{\bf T}_t^{\rm (EM)}$ is the 
total active stress-energy tensor of the source. The quantity $r_{\text s0}$ is 
thus the generalized Schwarzschild radius of the source at epoch $t_0$ 
modified with terms representing the electrostatic field energy. Note that, 
since isolated systems do not exist except as an approximation in 
quasi-metric gravity [3], the solution (41) is expected to be realistic only 
for ${\rho}{\ll}{\Xi}_0$. 

We may also ask how ${\bar B}({\rho})$ behaves for small distances in the case 
of a point source, i.e., in the limit where ${\rho}_{\rm sf}{\rightarrow}0$. 
Requiring correspondence with the uncharged case [3], we impose the condition 
${\bar B}({\rho}){\rightarrow}0$ in the limit ${\rho}{\rightarrow}C$, where 
$C$ is found from equation (41). The result is
\eqa
C={\frac{r_{\rm Q0}}{{\sqrt{{\rm artanh}^2
[{\frac{r_{\rm Q0}}{r_{\rm s0}}}]+{\frac{r_{\rm Q0}^2}{{\Xi}_0^2}}}}}}.
\ena
By taking the limit $r_{\rm Q0}{\rightarrow}0$ in equation (44) one finds the
right correspondence with the uncharged case. This means that the qualitative 
behaviour of the gravitational field does not depend on whether or not the 
source has a net charge.

To construct the metric family ${\bf g}_t$ from the family ${\bf {\bar g}}_t$ 
given in equation (31), one uses the method described in [1, 2]. (See also [3] 
for the case when $r_{\text Q0}$ vanishes.) To do this construction, we need the 
quantity $v({\rho})={\frac{{\bar B},_{\rho}}{2{\bar B}}}{\bar x}_{\cal F}^{\rho}c$.
We can calculate ${\bar x}_{\cal F}^{\rho}$ by introducing the Schwarzschild 
radial coordinate $r{\equiv}{\sqrt{{\bar B}}}{\rho}$ since
${\bf {\bar x}}_{\cal F}=r{\frac{\partial}{{\partial}r}}$ for all spherically 
symmetric cases [1, 2]. We then easily find
\eqa
{\bar x}_{\cal F}^{\rho}={\rho}{\Big [}1+{\frac{\rho}{2}}
{\frac{{\bar B},_{\rho}}{\bar B}}{\Big ]}^{-1}, \quad
{\Rightarrow} \quad v({\rho})={\frac{{\bar B},_{\rho}}{\bar B}}
{\Big [}{\frac{{\bar B},_{\rho}}{\bar B}}+{\frac{2}{\rho}}{\Big ]}^{-1}c,
\ena
and calculating the derivatives we find the exact expression
\eqa
v({\rho})={\frac{{\Big (}{\frac{r_{\rm s0}}{2{\rho}}}{\cosh}{\Big [}
{\frac{r_{\rm Q0}{\sqrt{1-{\frac{{\rho}^2}{{\Xi}_0^2}}}}}{\rho}}{\Big ]}
-{\frac{r_{\rm Q0}}{2{\rho}}}{\sinh}{\Big [}{\frac{r_{\rm Q0}
{\sqrt{1-{\frac{{\rho}^2}{{\Xi}_0^2}}}}}{\rho}}{\Big ]}{\Big )}c}
{{\Big (}{\sqrt{1-{\frac{{\rho}^2}{{\Xi}_0^2}}}}
+{\frac{r_{\rm s0}}{2{\rho}}}{\Big )}{\cosh}{\Big [}
{\frac{r_{\rm Q0}{\sqrt{1-{\frac{{\rho}^2}{{\Xi}_0^2}}}}}{{\rho}}}{\Big ]}
-{\Big (}{\frac{r_{\rm s0}}{r_{\rm Q0}}}
{\sqrt{1-{\frac{{\rho}^2}{{\Xi}_0^2}}}}+
{\frac{r_{\rm Q0}}{2{\rho}}}{\Big )}
{\sinh}{\Big [}{\frac{r_{\rm Q0}{\sqrt{1-
{\frac{{\rho}^2}{{\Xi}_0^2}}}}}{\rho}}{\Big ]}}}.
\ena
Then, using equations (2), (4) and (31), we obtain
\eqa
ds^2_t={\bar B}({\rho}){\Big \{}-{\Big (}1-{\frac{v^2({\rho})}{c^2}}{\Big )}^2
(dx^0)^2+({\frac{t}{t_0}})^2{\Big [}
{\Big (}{\frac{1+{\frac{v({\rho})}{c}}}{1-{\frac{v({\rho})}{c}}}}{\Big )}^2
{\frac{d{\rho}^2}{1-{\frac{{\rho}^2}{{\Xi}_0^2}}}}+
{\rho}^2d{\Omega}^2{\Big ]}{\Big \}}.
\ena
Substituting the above expressions for ${\bar B}({\rho})$ and $v({\rho})$ into 
equation (47) now yields an explicit, exact expression for the metric family 
${\bf g}_t$. Note that the qualitative behaviour of the exact solution (47) 
is quite similar to that found for the uncharged case presented in [3].

To compare the line element family (47) to its counterpart in GR, it will be 
useful to find a weak-field approximation using the Schwarzschild radial 
coordinate introduced above. To do that, we first notice that one may find an 
implicit expression for ${\bar B}(r)$ by substituting 
${\rho}={\frac{r}{\sqrt {{\bar B}}}}$ into equation (41). From this we 
see that ${\bar B}(r)$ cannot be written in closed form. However, the implicit 
expression thus found may be used to find a series solution for weak fields, 
in terms of the small quantities ${\frac{r_{\rm s0}}{r}}$, 
${\frac{r_{\rm Q0}}{r}}$ and ${\frac{r}{{\Xi}_0}}$. After some straightforward 
work, one may show that an approximate expression for ${\bar B}(r)$ is given by
\eqa
{\bar B}(r)=1-{\frac{r_{\text s0}}{r}}+{\Big (}1+
{\frac{r_{\text Q0}^2}{r_{\text s0}^2}}{\Big )}{\frac{r_{\text s0}^2}{2r^2}}+
{\frac{r_{\text s0}r}{2{\Xi}_0^2}}-{\Big (}1+
{\frac{22r_{\text Q0}^2}{3r_{\text s0}^2}}{\Big )}
{\frac{r_{\text s0}^3}{8r^3}}+{\cdots}.
\ena
We also define (with $'{\equiv}{\frac{{\partial}}{{\partial}r}}$)
\eqa
{\bar A}(r){\equiv}{\frac{{\Big [}1-{\frac{r}{2}}
{\frac{{\bar B}'(r)}{{\bar B}(r)}}
{\Big ]}^2}{1-{\frac{r^2}{{\bar B}(r){\Xi}^2_0}}}}=
1-{\frac{r_{\text s0}}{r}}+{\Big (}1+4{\frac{r_{\text Q0}^2}{r_{\text s0}^2}}
{\Big )}{\frac{r_{\text s0}^2}{4r^2}}+{\frac{r^2}{{\Xi}_0^2}}+{\cdots}.
\ena
The metric family ${\bf g}_t$ may now be represented in the form
\eqa
ds^2_t=-B(r)(dx^0)^2+({\frac{t}{t_0}})^2{\Big (}
A(r)dr^2+r^2d{\Omega}^2{\Big )},
\ena
where
\eqa
A(r){\equiv}{\Big (}{\frac{1+{\frac{v(r)}{c}}}{1-{\frac{v(r)}c}}}{\Big )}^2
{\bar A}(r), \qquad
B(r){\equiv}{\Big (}1-{\frac{v^2(r)}{c^2}}{\Big )}^2{\bar B}(r),
\ena
\eqa
v(r)={\frac{cr}{2}}{\frac{{\bar B}'(r)}{{\bar B}(r)}}=
c[{\frac{r_{\text s0}}{2r}}-{\frac{r_{\text Q0}^2}{2r^2}}+{\cdots}].
\ena
Inserting equations (48), (49) and (52) into equation (51) and equation (51) 
into equation (50), we finally find the wanted weak-field representation for 
${\bf g}_t$, i.e.,
\eqa
ds^2_t&=&-{\Bigg (}1-{\frac{r_{\text s0}}{r}}+{\frac{r_{\text Q0}^2}{2r^2}}+
{\frac{r_{\text s0}r}{2{\Xi}_0^2}}+{\Big (}1+
{\frac{2r_{\text Q0}^2}{9r_{\text s0}^2}}{\Big )}
{\frac{3r_{\text s0}^3}{8r^3}}+{\cdots}{\Bigg )}(dx^0)^2 \nonumber \\
&&+({\frac{t}{t_0}})^2{\Bigg (}{\Big \{}1+{\frac{r_{\text s0}}{r}}
+{\Big (}1-4{\frac{r_{\text Q0}^2}{r_{\text s0}^2}}{\Big )}
{\frac{r_{\text s0}^2}{4r^2}}+{\frac{r^2}{{\Xi}_0^2}}+
{\cdots}{\Big \}}dr^2+r^2d{\Omega}^2{\Bigg )}.
\ena
This expression represents the gravitational field (to the relevant accuracy)
outside a spherically symmetric, metrically static and charged source in 
quasi-metric gravity. Its correspondence with the Reissner-Nordstr\"{o}m
solution in GR may be found by setting ${\frac{t}{t_0}}$ equal to unity in
equation (53) and then taking the limit ${\Xi}_0{\rightarrow}{\infty}$
(with $G^{\rm B}=G^{\rm S}=G$). Note that an extension of the 
Reissner-Nordstr\"{o}m solution, adapted to a Friedmann-Robertson-Walker 
background, was found in [4]. But of course equation (53) does not have any 
further correspondence with this extended solution, since the cosmic expansion 
as described in the QMF is fundamentally different from its counterpart in the 
MF. That is, the way the cosmic expansion is fit into the extended 
Reissner-Nordstr\"{o}m solution has no counterpart in equation (53).

To calculate the paths of charged test particles, one uses the quasi-metric
equations of motion (14), (15). But to apply these equations, one needs to know 
the Lorentz force from equation (18). This means that we must first calculate 
${\bf {\bar F}}_t$ from equation (19) and then find ${\bf F}_t$ from $v$ and
${\bf {\bar F}}_t$. That is, using equation (19) we first find the passive 
electric field ${\bf {\bar E}}_t$:
\eqa
{\bar E}_{(t){\rho}}{\equiv}{\bar F}_{(t){\bar {\perp}}{\rho}}
={\frac{t_0}{t}}{\bar N}_t^{-1}{\tilde F}_{(t){\bar {\perp}}{\rho}}
={\frac{t_0}{t}}{\frac{Q}{{\rho}^2
{\sqrt{{\bar B}({\rho})[1-{\frac{{\rho}^2}{{\Xi}_0^2}}]}}}}, \quad
{\bar E}_{(t)r}{\equiv}{\bar F}_{(t){\bar {\perp}}r}
={\frac{t_0}{t}}{\frac{{\sqrt{{\bar A}(r)}}Q}{r^2}}.
\ena
Note that, since ${\bar A}(r)$ and ${\bar B}(r)$ are not inverse functions,
${\bar F}_{(t_0)r0}={\sqrt{{\bar B}(r)}}{\bar E}_{(t_0)r}$ does not take the 
Euclidean form. On the other hand, for ${\frac{{\rho}^2}{{\Xi}_0^2}}{\ll}1$,
${\bar F}_{(t_0){\rho}0}={\sqrt{{\bar B}({\rho})}}{\bar E}_{(t_0){\rho}}$ 
approximately does. Second, we are able to find $F_{(t)0r}=
(1+{\frac{v(r)}{c}})^2{\bar F}_{(t)0r}$ (and a similar formula for 
$F_{(t)0{\rho}}$) using equation (3), and we then get (using equation (54))
\eqa
E_{(t){\rho}}{\equiv}F_{(t){\perp}{\rho}}
={\Big (}{\frac{1+{\frac{v({\rho})}{c}}}{1-{\frac{v({\rho})}{c}}}}{\Big )}
{\bar E}_{(t){\rho}}, \quad E_{(t)r}{\equiv}F_{(t){\perp}r}
={\Big (}{\frac{1+{\frac{v(r)}{c}}}{1-{\frac{v(r)}{c}}}}{\Big )}
{\bar F}_{(t){\bar {\perp}}r}={\frac{t_0}{t}}{\frac{{\sqrt{A(r)}}Q}{r^2}}.
\ena
We may now calculate the passive mass-energy density 
${\varrho}_{\text m}^{\rm (EM)}$ and pressure $p^{\rm (EM)}$ of the electromagnetic 
field as measured in the metrically static frame. This yields
\eqa
{\bar {\cal T}}^{\rm (EM)}_{(t){\bar {\perp}}{\bar {\perp}}}=
{\varrho}_{\text m}^{\rm (EM)}c^2=
{\frac{1}{8{\pi}}}{\bar E}_{(t){\rho}}{\hat {\bar E}}_{(t)}^{\rho}=
{\frac{1}{8{\pi}}}{\bar E}_{(t)r}{\hat {\bar E}}_{(t)}^r=
{\frac{1}{8{\pi}}}E_{(t)r}{\hat E}_{(t)}^r=
{\frac{t_0^4}{t^4}}{\frac{Q^2}{8{\pi}r^4}},
\ena
\eqa
{\bar {\cal T}}_{(t){\theta}}^{{\rm (EM)}{\theta}}=
{\bar {\cal T}}_{(t){\phi}}^{{\rm (EM)}{\phi}}=
-{\bar {\cal T}}_{(t)r}^{{\rm (EM)}r}=
-{\bar {\cal T}}_{(t){\rho}}^{{\rm (EM)}{\rho}}=
p^{\rm (EM)}={\bar {\cal T}}^{\rm (EM)}_{(t){\bar {\perp}}{\bar {\perp}}}.
\ena
On the other hand, we can define the passive stress-energy tensor 
${\cal T}^{\rm (EM)}_t$ of the electromagnetic field in $({\cal N},{\bf g}_t)$
from ${\bf F}_t$ by
\eqa
{\cal T}_{(t){\alpha}{\beta}}^{\rm (EM)}={\frac{1}{4{\pi}}}{\Big (}
F_{(t){\alpha}}^{{\ }{\ }{\ }{\,}{\,}{\nu}}
F_{(t){\beta}{\nu}}-{\frac{1}{4}}F_{(t){\sigma}{\rho}}
F_{(t)}^{{\sigma}{\rho}}g_{(t){\alpha}{\beta}}{\Big )}
{\sqrt{{\frac{{\bar h}_t}{h_t}}}},
\ena
where the square root factor is necessary to ensure that the total amount of 
passive electromagnetic mass-energy is identical in 
$({\cal N},{\bf {\bar g}}_t)$ and $({\cal N},{\bf g}_t)$. Using equation (58)
we may then calculate
\eqa
{\cal T}^{\rm (EM)}_{(t){\perp}{\perp}}=
{\frac{1}{8{\pi}}}E_{(t)r}{\hat E}_{(t)}^r{\sqrt{\frac{{\bar h}_t}{h_t}}}
={\varrho}_{\text m}^{\rm (EM)}c^2{\sqrt{\frac{{\bar h}_t}{h_t}}}=
{\frac{t_0^4}{t^4}}{\frac{Q^2}{8{\pi}r^4}}
{\Big (}{\frac{1-{\frac{v(r)}{c}}}{1+{\frac{v(r)}{c}}}}{\Big )},
\ena
and similar formulae for the other components of ${\cal T}_t^{\rm (EM)}$.
\subsection{The effects of cosmic expansion on 
electromagnetism}
To see to what extent a classical, electromagnetically bound system is 
affected by the global cosmic expansion, we may calculate the path of a charged
test particle with mass $m$ and charge $q$ in the electric field of an 
isolated spherical charge $Q$. That is, we may use the metric family (50) where 
it is assumed that the gravitational field of the source is negligible, i.e., 
$B(r)=1$. We also set $A(r)=1$ since isolated systems are realistic in the QMF 
only if $r{\ll}{\Xi}_0$ [3]. Using equation (55) and the coordinate expression 
for ${\bf u}_t$, i.e., $u_{(t)}^{\alpha}{\equiv}{\frac{dx^{\alpha}}{d{\tau}_t}}$ 
where ${\tau}_t$ is the proper time as measured along the path of the test 
particle, equation (18) then yields the components of the Lorentz force 
(neglecting radiative effects and taking care of the fact that $qQ<0$ since we 
have a bound system)
\eqa
ma_{(t)}^r=-{\frac{t_0^3}{t^3}}
{\frac{{\mid}qQ{\mid}}{r^2}}{\frac{dx^0}{cd{\tau}_t}}, \qquad
ma_{(t)}^0=-{\frac{t_0}{t}}{\frac{{\mid}qQ{\mid}}{r^2}}{\frac{dr}{cd{\tau}_t}}.
\ena
Furthermore, we may insert this result into the equations of motion (14), (15). 
Confining the motion of the test particle to the equatorial plane 
${\theta}={\pi}/2$, the result is 
\eqa
{\frac{d^2r}{d{\tau}_t^2}}-r{\Big (}{\frac{d{\phi}}{d{\tau}_t}}{\Big )}^2
+{\frac{1}{ct}}{\frac{dr}{d{\tau}_t}}{\frac{dx^0}{d{\tau}_t}}=
-{\frac{t_0^3}{t^3}}{\frac{{\mid}qQ{\mid}}{mr^2}}{\frac{dx^0}{cd{\tau}_t}},
\ena
\eqa
{\frac{d^2x^0}{d{\tau}_t^2}}=-
{\frac{t_0}{t}}{\frac{{\mid}qQ{\mid}}{mr^2}}{\frac{dr}{cd{\tau}_t}},
\ena
\eqa
{\frac{d^2{\phi}}{d{\tau}_t^2}}+{\frac{2}{r}}{\frac{d{\phi}}{d{\tau}_t}}
{\frac{dr}{d{\tau}_t}}+
{\frac{1}{ct}}{\frac{d{\phi}}{d{\tau}_t}}{\frac{dx^0}{d{\tau}_t}}=0.
\ena
Equation (63) yields a constant of motion $J$, namely [3]
\eqa
J{\equiv}{\frac{t}{t_0}}r^2{\frac{d{\phi}}{cd{\tau}_t}}.
\ena
Introducing the dimensionless variable ${\xi}{\equiv}{\frac{t}{t_0}}$ and 
neglecting terms of order ${\frac{{\xi}^2}{{\Xi}_0^2}}$, equations (60)-(64)
may be combined and written in the form (since $dx^0=cdt$ along the path of 
the charged particle)
\eqa
(1+{\frac{J^2}{r^2}}){\Big [}{\frac{d^2r}{d{\xi}^2}}+{\frac{1}{\xi}}
{\frac{dr}{d{\xi}}}{\Big ]}+{\frac{{\mid}qQ{\mid}}{{\xi}mcr^2}}
{\sqrt{1+{\frac{J^2}{r^2}}}}{\Big [}{\frac{{\Xi}_0^2}{{\xi}^2}}-
({\frac{dr}{d{\xi}}})^2{\Big ]}-{\frac{{\Xi}_0^2J^2}{{\xi}^2r^3}}=0,
\ena
\eqa
{\frac{dt}{d{\tau}_t}}={\sqrt{1+{\frac{J^2}{r^2}}}},
\ena
where we have used equation (50) (with $B(r)=A(r)=1$). Equation (65) is too
complicated to have any hope of finding an exact solution. However, in the
non-relativistic limit, one may neglect the term quadratic in the radial 
velocity. Then, taking as an initial condition that ${\frac{dr}{d{\xi}}}$
should vanish, numerical calculations show that the solution lies close to 
(oscillates around) a straight line $r({\xi}){\propto}{\xi}$. This seems 
reasonable, since one expects that the Lorentz force term and the centrifugal 
term in equation (65) should on average scale similarly with respect to the 
factor ${\frac{t}{t_0}}$. 

Since the solution of equation (65) yields that $r(t){\propto}{\frac{t}{t_0}}$ 
on average, we have that according to the QMF, a classical system bound 
solely by electromagnetic forces will experience an even larger secular
expansion than a gravitationally bound system of the same size. However, an 
important difference is that the norm of the (passive) electromagnetic field
is constant at a given constant (proper) distance from the source (as seen from 
equation (55)), whereas the gravitational field at a given constant distance 
from the source gets stronger due to the secular increase of active 
mass-energy [3]. (This last point can also easily be seen from equation (53).)

This means that the effect of the cosmic expansion on the electromagnetic 
field is fundamentally different from its effect on the gravitational field. 
In particular it means that, except for a global cosmic attenuation not 
noticeable locally, the electromagnetic field is unaffected by the global 
cosmic expansion, unlike the gravitational field. And even if it is predicted 
that a classical, electromagnetically bound system should experience cosmic 
expansion, this expansion implies that the potential energy of the system 
increases and that its kinetic energy decreases. But suitable interactions 
with other systems should then allow the system to return to its initial 
state. On the other hand, it is impossible to null out the effects of the 
cosmic expansion on a gravitational system via interactions with other 
systems [3].

For ``small'' classical systems, radiative effects will dominate over the 
expansion effect by many orders of magnitude so the expansion would not be 
noticeable. Furthermore, the classical calculation is hardly relevant for
quantum-mechanical systems since quantum-mechanical states should not be 
affected by the cosmic expansion. This is readily seen for the Bohr atom, 
where the radii of the ``permitted'' electron orbits are determined from the 
requirement that angular momenta take the discrete values $L_{\rm n}=n{\hbar}$;
i.e., a set of static values. But this is inconsistent with the above solution 
of the classical system yielding $L{\propto}{\frac{t}{t_0}}$, which is
non-static. This means that there is no reason to expect that 
quantum-mechanical systems such as atoms should behave as purely classical 
systems and increase in size due to the global cosmic expansion. (Rather 
the expansion could perhaps induce spontaneous excitations. To explore this 
possibility, a quantum-mechanical calculation should be carried out.) For a 
comparison with GR, see references [5] and [6] for the estimated effects of 
the cosmic expansion on a classical ``atom'' according to GR. Note that the 
results found in these papers differ considerably from the results found here.
\section{Geometric optics in quasi-metric space-time}
In this section, we sketch how the fundamental laws of geometric optics are
derived within the quasi-metric framework. Except for small changes, these
derivations may be done exactly as for the metric framework, see, e.g., 
reference [7]. 

The main difference from the metric framework, is that we are forced to include
the effects of the cosmic expansion on the wavelength, the amplitude and the
polarization of electromagnetic waves propagating through a source-free
region of space-time. But the electromagnetic waves may still be taken to
be locally monochromatic and plane-fronted. 

Similarly to the metric case, we start with a vector potential family
${\bf {\bar A}}_t$ of the form
\eqa
{\bar A}_{(t)}^{\mu}=
{\Re}[{\mid}{\bar {\cal A}}_t{\mid}{\exp}(i{\bar {\vartheta}}_t)
{\bar f}_{(t)}^{\mu}], \qquad {\bar f}_{(t)}^{\mu}{\equiv}
{\mid}{\bar {\cal A}}_t{\mid}^{-1}{\bar {\cal A}}_{(t)}^{\mu},
\ena 
where ${\mid}{\bf {\bar {\cal A}}}_t{\mid}$ denotes the norm of the (possibly
complex) vector amplitude ${\bar {\cal A}}_{(t)}^{\mu}
{\frac{\partial}{{\partial}{x^{\mu}}}}$ of ${\bf {\bar A}}_t$ and where 
${\bf {\bar f}}_t$ is the polarization vector family. Moreover, the phase
factor ${\bar {\vartheta}}_t$ is defined by (using a GTCS)
\eqa
{\bar {\vartheta}}_t{\equiv}{\bar k}_{(t)0}(x^0-x^0_1)+{\bar k}_{(t)s}x^s, 
\qquad {\bar k}_{(t){\mu}}={\bar {\vartheta}}_{t},_{\mu},
\ena
where ${\bf {\bar k}}_t$ is the wave vector family and $x^0_1=ct_1$ is an 
arbitrary reference epoch. Note that equations (67) and (68) are identical 
to those valid for the metric case except for the dependence on $t$. The 
$t$-dependence of ${\bf {\bar k}}_t$ is determined by how it is affected by 
the global cosmic expansion. That is, in the QMF there is a general 
cosmological attenuation of the electromagnetic field, and for electromagnetic 
radiation this attenuation takes the form of a general redshift due to the 
global cosmic expansion. This means that we must have (using a GTCS)
\eqa
{\bar k}_{(t)0},_t&=&-{\frac{1}{t}}{\bar k}_{(t)0}, \quad 
{\bar k}_{(t)j},_t={\frac{1}{2}}{\bar k}_{(t)s}{\tilde h}_{(t)}^{ks}
{\tilde h}_{(t)kj},_t, \nonumber \\
{\bar k}_{(t)}^0,_t&=&-({\frac{1}{t}}+
2{\frac{{\bar N}_{t,t}}{{\bar N}_t}}){\bar k}_{(t)}^0, \quad 
{\bar k}_{(t)}^j,_t=-2({\frac{1}{t}}+{\frac{{\bar N}_{t,t}}{{\bar N}_t}})
{\bar k}_{(t)}^j-{\frac{1}{2}}{\bar k}_{(t)}^s
{\tilde h}_{(t)}^{kj}{\tilde h}_{(t)ks},_t.
\ena
The $t$-dependence of ${\mid}{\bar {\cal A}}_t{\mid}$ follows from equations
(17) and (67), i.e., ${\mid}{\bar {\cal A}}_t{\mid},_t
=-({\frac{1}{t}}+{\frac{{\bar N},_t}{{\bar N}_t}}){\mid}{\bar {\cal A}}_t
{\mid}$. Together with the $t$-dependences given in equation (69), this
then implies that the covariant derivatives of ${\bf {\bar k}}_t$ and of 
${\bf {\bar f}}_t$ in the $t$-direction are given by (see [1, 2] for the 
connection coefficients)
\eqa
{\stackrel{\star}{\bf {\bar{\nabla}}}}_{\frac{\partial}{{\partial}t}}
{\bf {\bar k}}_t=-{\Big (}{\frac{1}{t}}+{\frac{{\bar N}_t,_t}{{\bar N}_t}}
{\Big )}{\bf {\bar k}}_t, \qquad
{\stackrel{\star}{\bf {\bar{\nabla}}}}_{\frac{\partial}{{\partial}t}}
{\bf {\bar f}}_t=-i{\frac{{\partial}{\bar {\vartheta}}_t}{{\partial}t}}
{\bf {\bar f}}_t, \qquad
{\bf {\stackrel{\star}{\bar {\nabla}}}}_{\frac{\partial}{{\partial}t}}
({\frac{t^3}{t_0^3}}{\bar N}_t^3{\mid}{\bar {\cal A}}_t{\mid}^2
{\bf {\bar k}}_t)=0.
\ena
We now put equation (67) into the Lorentz gauge condition 
${\bar A}_{(t);{\mu}}^{\mu}=0$ and then into Maxwell's equations (24) without 
sources. These calculations are done explicitly for the metric case in 
reference [7]. Since the derivations for the quasi-metric case are very 
similar, we will not repeat them here. Rather we list the results, also very 
similar to those found for the metric case. Firstly, we find that
\eqa
{\bar k}_{(t){\mu}}{\bar f}_{(t)}^{\mu}=0, \qquad 
{\bar k}_{(t){\mu}}{\bar k}_{(t)}^{\mu}=0, \qquad
{\bar k}_{(t){\mu};{\nu}}{\bar k}_{(t)}^{\nu}=0.
\ena
Selecting a suitable affine parameter ${\lambda}$ along a light ray, we also
have that
\eqa
{\bf {\stackrel{\star}{\bar {\nabla}}}}_{\frac{\partial}{{\partial}{\lambda}}}
{\bf {\bar k}}_t{\equiv}{\frac{dt}{d{\lambda}}}
{\bf {\stackrel{\star}{\bar {\nabla}}}}_{\frac{\partial}{{\partial}t}}
{\bf {\bar k}}_t+{\frac{dx^{\alpha}}{d{\lambda}}}
{\bf {\stackrel{\star}{\bar {\nabla}}}}_
{\frac{\partial}{{\partial}x^{\alpha}}}{\bf {\bar k}}_t.
\ena
Equations (70), (71) and (72) then yield (in component notation, using a GTCS)
\eqa
{\frac{d{\bar k}_{(t)}^{\mu}}{d{\lambda}}}+{\Big (}
{\topstar{\bar {\Gamma}}}^{\mu}_{t{\epsilon}}{\frac{dt}{d{\lambda}}}+
{\topstar{\bar {\Gamma}}}^{\mu}_{{\nu}{\epsilon}}{\frac{dx^{\nu}}{d{\lambda}}}
{\Big )}{\bar k}_{(t)}^{\epsilon}=
-{\Big (}{\frac{1}{t}}+{\frac{{\bar N}_t,_t}{{\bar N}_t}}{\Big )}
{\frac{dt}{d{\lambda}}}{\bar k}_{(t)}^{\mu}.
\ena
Provided that we make the identification
\eqa
{\frac{dx^{\mu}}{d{\lambda}}}={\frac{t}{t_0}}{\bar k}_{(t)}^{\mu},
\ena
between the tangent vector field along the light ray and ${\bf {\bar k}}_t$,
and if one makes a suitable change of parameter along the 
path for the case when ${\bar N}_t$ depends on $t$, equation (73) is identical
to the geodesic equation in $({\cal N},{\bf {\bar g}}_t)$. That is, we have 
derived from Maxwell's equations that light rays are null geodesics in 
$({\cal N},{\bf {\bar g}}_t)$. But since any ``physical'' null vector is
required to remain a null vector under the transformation 
${\bf {\bar g}}_t{\rightarrow}{\bf g}_t$, we may write 
${\bf {\bar k}}_t{\rightarrow}{\bf k}_t$, where ${\bf {\bar k}}_t$ transforms
as shown in equation (1). Similarly, since the norm of ${\bf {\bar f}}_t$ is 
required to be constant under the transformation 
${\bf {\bar g}}_t{\rightarrow}{\bf g}_t$, we may write ${\bf {\bar f}}_t
{\rightarrow}{\bf f}_t$, where ${\bf {\bar f}}_t$ transforms in the same way
as ${\bf {\bar k}}_t$. Furthermore, we may define a new phase factor
${\vartheta}_t$ by
\eqa
{\vartheta}_t{\equiv}k_{(t)0}(x^0-x^0_1)+k_{(t)s}x^s, 
\qquad k_{(t){\mu}}={\vartheta}_{t},_{\mu},
\ena
where the $t$-dependence of ${\bf k}_t$ is given by (in a GTCS)
\eqa
k_{(t)0},_t&=&-{\frac{1}{t}}k_{(t)0}, \qquad k_{(t)j},_t={\frac{1}{2}}
k_{(t)s}{\hat h}_{(t)}^{ks}{\hat h}_{(t)kj},_t, \nonumber \\
k_{(t)}^0,_t&=&-{\frac{1}{t}}k_{(t)}^0, \qquad 
k_{(t)}^j,_t=-{\frac{2}{t}}k_{(t)}^j-{\frac{1}{2}}
k_{(t)}^s{\hat h}_{(t)}^{kj}{\hat h}_{(t)ks},_t.
\ena
From equations (75), (76) and the definitions of ${\bf k}_t$ and ${\bf f}_t$,
we get the counterpart in $({\cal N},{\bf g}_t)$ to equation (71), namely
\eqa
k_{(t){\mu}}f_{(t)}^{\mu}=0, \qquad k_{(t){\mu}}k_{(t)}^{\mu}=0, \qquad
k_{(t){\mu};{\nu}}k_{(t)}^{\nu}=0.
\ena
Using the counterparts in $({\cal N},{\bf g}_t)$ to equations (70), (72) and
(74) together with equations (76) and (77), it is now straightforward to derive
the counterpart in $({\cal N},{\bf g}_t)$ to equation (73). Thus we have shown
that light rays must be null geodesics in $({\cal N},{\bf g}_t)$ as well. 

Secondly, we find that just as for the metric case, we have
\eqa
{\bar {\cal A}}_{(t);{\nu}}^{\mu}{\bar k}_{(t)}^{\nu}=-{\frac{1}{2}}
{\bar k}_{(t);{\alpha}}^{\alpha}{\bar {\cal A}}_{(t)}^{\mu}, \quad 
\Rightarrow \quad
{\bf {\stackrel{\star}{\bar {\nabla}}}}_{{\bf {\bar k}}_t}{\bf {\bar f}}_t
{\equiv}c^{-1}{\bar k}_{(t)}^0
{\bf {\stackrel{\star}{\bar {\nabla}}}}_{\frac{\partial}{{\partial}t}}
{\bf {\bar f}}_t+{\bar k}_{(t)}^{\nu}
{\bf {\stackrel{\star}{\bar {\nabla}}}}_{\frac{\partial}{{\partial}x^{\nu}}}
{\bf {\bar f}}_t=-{\frac{i}{c}}{\bar k}_{(t)}^0
{\frac{{\partial}{\bar {\vartheta}}_t}{{\partial}t}}{\bf {\bar f}}_t,
\ena
where we have used equation (70) and equation (79) below in the last step. 
That is, the polarization vector family is perpendicular to the light rays and 
parallel-transported along them. From equation (77) we see that the 
polarization vector family is perpendicular to the light rays in 
$({\cal N},{\bf g}_t)$ as well. Besides, since light rays are also null 
geodesics in $({\cal N},{\bf g}_t)$, ${\bf f}_t$ is parallel-transported along 
them and it follows that ${\bf {\stackrel{\star}{\nabla}}}_{{\bf k}_t}{\bf f}_t$ 
must be proportional to ${\bf f}_t$.

Thirdly, also just as for the metric case, we have that
\eqa
({\mid}{\bar {\cal A}}_t{\mid}^2{\bar k}_{(t)}^{\alpha})_{;{\alpha}}=0, 
\ena
or equivalently, that the volume integral ${\int}{\int}{\int}{\mid}
{\bar {\cal A}}_t{\mid}^2{\bar k}_{(t){\bar {\perp}}}{\sqrt{{\bar h}_t}}d^3x$ 
has a constant value when integrating over the 3-volume cut out of the FHSs by
a tube formed of light rays (such that no rays cross the tube walls). This is
the law of conservation of photon number in geometric optics. And since
${\bar k}_{(t){\bar {\perp}}}=k_{(t){\perp}}$, the above integral may be 
written as ${\int}{\int}{\int}{\mid}{\cal A}_t{\mid}^2k_{(t){\perp}}
{\sqrt{h}_t}d^3x$ where ${\mid}{\cal A}_t{\mid}^2{\sqrt{h_t}}{\equiv}
{\mid}{\bar {\cal A}}_t{\mid}^2{\sqrt{{\bar h}_t}}$. So we see that the law of
conservation of photon number holds for $({\cal N},{\bf g}_t)$ also but with a
different ``density of light rays'' ${\mid}{\cal A}_t{\mid}^2k_{(t){\perp}}$ 
on each FHS. However, the total number of light rays in the tube is the same.

The passive electromagnetic field tensor ${\bf {\bar F}}_t$ for light rays
has the same form as for the metric case, i.e.,
\eqa
{\bar F}_{(t){\mu}{\nu}}={\Re}[i{\mid}{\bar {\cal A}}_t{\mid}
{\exp}(i{\bar {\vartheta}}_t)({\bar k}_{(t){\mu}}{\bar f}_{(t){\nu}}-
{\bar f}_{(t){\mu}}{\bar k}_{(t){\nu}})].
\ena
We may find ${\bf F}_t$ from ${\bf {\bar F}}_t$ just as ${\bf g}_t$ is
found from ${\bf {\bar g}}_t$ (see equations (2)-(4)). Moreover, we may also
include the active aspects of light rays as long as we are not going beyond 
the geometric optics approximation. That is, we can construct 
${\bf {\tilde F}}_t$ from ${\bf {\bar F}}_t$ using equation (19). The
passive electromagnetic field stress-energy tensor 
${\bf {\bar {\cal T}}}_t^{\rm (EM)}$ then yields the active electromagnetic field
stress-energy tensor ${\bf T}_t^{\rm (EM)}$ via equation (22). Thus, putting 
equation (80) into the standard definition of 
${\bf {\bar {\cal T}}}_t^{\rm (EM)}$, we get the expressions (averaged over one 
wavelength)
\eqa
{\bar {\cal T}}_{(t){\mu}{\nu}}^{\rm (EM)}={\frac{1}{8{\pi}}}{\mid}
{\bar {\cal A}}_t{\mid}^2{\bar k}_{(t){\mu}}{\bar k}_{(t){\nu}}, \qquad 
\Rightarrow \qquad  T_{(t){\mu}{\nu}}^{\rm (EM)}=
{\frac{t^2}{t_0^2}}{\bar N}_t^2{\frac{1}{8{\pi}}}
{\mid}{\bar {\cal A}}_{t}{\mid}^2{\bar k}_{(t){\mu}}{\bar k}_{(t){\nu}}.
\ena
This expression for ${\bf T}_t^{\rm (EM)}$ can then be used to find the necessary
projections to be inserted into the gravitational field equations. So as long 
as the geometric optics approximation holds, it is thus possible to set up a 
well-defined initial value problem for light rays coupled to gravity within 
the QMF.
\section{Conclusion}
In metric theory, the nature of the cosmic expansion is kinematical (in the
general sense of the word). That is, to which degree a given system is 
influenced by the cosmic expansion is determined by dynamical laws subject to 
cosmological initial conditions. This means that in metric theory, bound 
systems may in principle be influenced by the cosmic expansion regardless of 
the nature of the force holding the system together. However, calculations 
show that the effect of the cosmic expansion on realistic local systems should 
be totally negligible; see, e.g., [5] and references listed therein. On the 
other hand, in the QMF, the nature of the cosmic expansion is non-kinematical, 
i.e., the expansion is not part of space-time's causal structure and it is 
described as a secular global change of scale between gravitational and 
non-gravitational systems. That is, in the QMF, the effects of the global 
cosmic expansion on gravitationally bound objects should be fundamentally 
different from its effects on objects solely bound by non-gravitational 
forces, e.g., electromagnetism.

In this paper, we have shown how to formulate classical electrodynamics coupled 
to gravity in a way consistent with the QMF. This is possible provided that 
the main effect of the cosmic expansion on the electromagnetic field takes 
the form of a global cosmic attenuation not noticeable locally. Moreover, as 
an illustrative example, we have calculated the exact solutions for the 
electric and gravitational fields in the electrovacuum outside a spherically 
symmetric, metrically static, charged source. This example shows that the 
effect of the cosmic expansion on the electric field is fundamentally 
different from its effect on the gravitational field. Also calculated is the 
path of a charged test particle moving in the electric field of (and 
electromagnetically bound to) an isolated spherical charge with negligible 
gravity. This calculation shows that the cosmic expansion affects the 
electromagnetically bound system even more than a similar, but gravitationally 
bound system. However, since the cosmic expansion should not affect quantum 
states, there is no reason to think that the calculated expansion of classical 
electromagnetically bound systems should apply to quantum-mechanical systems.

Furthermore we have shown that, just as for metric theory, within the QMF it 
is possible to derive the fact that light rays move along geodesics from 
Maxwell's equations in curved space-time. Also valid within the QMF are the 
other two main results of geometric optics in curved space-time. It thus 
seems that the QMF represents a self-consistent framework within which to do 
electromagnetism as well as relativistic gravitation.
\\ [4mm]
{\bf Acknowledgment} \\ [1mm]
I wish to thank Dr. K{\aa}re Olaussen for making a review of the manuscript.
\\ [4mm]
{\bf References} \\ [1mm]
{\bf [1]} D. {\O}stvang, {\em Gravit. Cosmol.} {\bf 11}, 205 (2005)
(gr-qc/0112025). \\
{\bf [2]} D. {\O}stvang, {\em Doctoral Thesis} (2001) (gr-qc/0111110). \\
{\bf [3]} D. {\O}stvang, {\em Gravit. Cosmol.} {\bf 13}, 1 (2007)
(gr-qc/0201097). \\
{\bf [4]} C.J. Gao, S.N. Zhang, {\em Phys. Lett.} {\bf B 595}, 28 (2004). \\
{\bf [5]} W.B. Bonnor, {\em Class. Quantum Grav.} {\bf 16}, 1313 (1999). \\
{\bf [6]} R.H. Price, gr-qc/0508052 (2005). \\
{\bf [7]} C.W. Misner, K.S. Thorne, J.A. Wheeler,
{\em Gravitation}, W.H. Freeman ${\&}$ Co. (1973). 
\end{document}